\documentstyle[sprocl]{article}

\input{psfig}
\def\PRL{\em Phys. Rev. Lett.}
\def\PRD{{\em Phys. Rev.} D}
\def\ZPC{{\em Z. Phys.} C}
\def\be{\begin{equation}}
\def\ee{\end{equation}}
\def\bea{\begin{eqnarray}}
\def\eea{\end{eqnarray}}
\begin{document}

\def\zpp{$0^{++}$}
\def\fz{$f_0(980)$}
\def\az{$a_0(980)$}
\def\Kz{$K_0^*(1430)$}
\def\fzz{$f_0(1300)$}
\def\fzzz{$f_0(1200-1300)$}
\def\azz{$a_0(1450)$}
\def\ss{$ s\bar s $}
\def\uu{$u\bar u+d\bar d$}
\def\qq{$q\bar q$}
\def\KK{$K\bar K$}
\def\sig{$\sigma$}
\def\lsim{\;\raise0.3ex\hbox{$<$\kern-0.75em\raise-1.1ex\hbox{$\sim$}}\;}
\def\gsim{\raise0.3ex\hbox{$>$\kern-0.75em\raise-1.1ex\hbox{$\sim$}}}
%\newcommand{\be}{\begin{eqnarray}}
%\newcommand{\ee}{\end{eqnarray}}

%\setcounter{secnumdepth}{2} % Number sections and subsections

%%%%%%%%%%%%%%%%%%%%%%%%%%%%%%%%%%%%%%%%%%%%%%%%%%
%                                                %
%    BEGINNING OF TEXT                           %
%                                                %
%%%%%%%%%%%%%%%%%%%%%%%%%%%%%%%%%%%%%%%%%%%%%%%%%%
%\pagestyle{empty}
\hfill  HU-SEFT R 1996-18

\title{THE SCALAR $q\bar q$ NONET AND \\ CONFIRMATION OF THE
BROAD  $\sigma(\approx 500)$ MESON\footnote{Invited talk at the International
conference: 
"Quark Confinement and the Hadron Spectrum II", June 26-29 1996, Como, Italy }}

\author{Nils A. T\"ornqvist }
\address
{University of Helsinki, SEFT, P.O. Box 9, Fin-00014 Helsinki, Finland; \\
email: Tornqvist@phcu.Helsinki.Fi}


\maketitle\abstracts{The 
available data on the $a_0(980)$, $f_0(980)$,
$f_0(1300)$ and $K^*_0(1430)$ mesons are fitted 
as a distorted $0^{++}$ \qq\ nonet using 
only 6 parameters and a very general model. This
includes all light two-pseudoscalar  thresholds, constraints from Adler
zeroes, flavour symmetric couplings, 
unitarity and physically acceptable analyticity. 
One finds that with the large overall
coupling there can appear  two physical resonance poles from only 
one $q\bar q$
state. Thus  the \fz\ and \fzz\ resonance poles are two manifestations of
the same \ss\ state. 
On the other hand, the \uu\ state, when unitarized and strongly distorted 
by hadronic mass shifts,
becomes an extremely broad Breit-Wigner-like background,
$m_{BW}=860$ MeV, $\Gamma_{BW}=880$ MeV, with its
pole at $s$=(0.158-i0.235) GeV$^2$. This is the \sig\ meson
required by models for spontaneous breaking of chiral symmetry.
}

\section{ Introduction}
This paper is a short summary of two recent papers~\cite{NAT,NAT2}, 
including a few new comments. These results 
give a new understanding of the controversial light scalar
mesons. It is  shown that one can describe the S-wave data on the
light   \qq\ nonet with   a model which includes most well established
theoretical constraints: Adler zeroes as required by chiral symmetry,
all light two-pseudoscalar (PP) thresholds with flavor symmetric couplings,
physically acceptable analyticity, and unitarity.  A unique
feature of this model is that it simultaneously describes the  whole scalar
nonet and one obtains a good representation of a large
set of relevant data. Only six parameters,
which all have a  clear physical interpretation, are needed: 
an overall coupling constant
($\gamma=1.14$), the bare mass of the $u\bar u$ or $d\bar d$ state
($m_0=1.42$ GeV), the extra mass for a strange quark ($m_s-m_u=100$ MeV),
a cutoff parameter ($k_0=0.56$ GeV/c), an Adler zero parameter for $K\pi$
($s_{A_{K\pi}}=-0.42$ GeV), and a phenomenological 
parameter enhancing the $\eta\eta '$ couplings ($\beta=1.6$). 

\section{ Understanding the S-waves}

%\newpage

In Figs.~1-3 we show the obtained fits to the $K\pi$, $\pi\pi$ S-waves and
to the  \az\ resonance peak in $\pi\eta$. The partial 
wave amplitude is in the case of one 
\qq\ resonance, such as the \az\ can be written 
\be
A(s)=-Im\Pi_{\pi\eta}(s)/[m_0^2+Re\Pi (s)-s +iIm\Pi (s)], \label{PWA}
\ee
\noindent where
\bea
Im\Pi (s)&=&\sum_i Im\Pi_i(s)  \label{impi}\\
       &=&-\sum_i \gamma_i^2(s-s_{A,i})\frac{k_i}{\sqrt s}e^{-k_i^2/k_0^2}
           \theta(s-s_{th,i})\ ,\nonumber \\
Re\Pi_i(s)&=&\frac 1\pi{\cal P}\int^\infty_{s_{th,1}} \frac{Im\Pi_i (s)}{s'-s} ds'
\ . \label{repi}
\eea

Here the coupling constants $\gamma_i$ are related by flavour
symmetry and the OZI rule, such that there is only one over all parameter
$\gamma$. The $s_{A,i}$ are the positions of the Adler
zeroes, which normally are  $s_{A,i}=0$, except  $s_{A,\pi\pi}=m_\pi^2/2$, and  
$s_{A,K\pi}$, which is a free parameter.
%\newpage 

 In the flavourless channels 
the situation is a little more complicated than eqs. (1-3)
since one has both \uu\ and \ss\  states, requiring a two dimensional 
mass matrix (See Ref.~\cite{NAT}).
 Note that the sum runs over all light
PP thresholds, which means three for the \az : $\pi\eta ,\ K\bar K,\pi\eta'$
and three for the 
\Kz : $ K\pi ,\ K\eta ,\ K\eta' $, while for the $f_0$'s there are
five channels: $\pi\pi ,K\bar K ,\ \eta\eta ,\ \eta\eta' ,\ \eta'\eta'$.
Five channels means the amplitudes have $2^5=32$! different Riemann sheets, 
and in principle there can be poles on each of these sheets.
In Fig.~4 we show as an example the running mass, $m_0^2+Re\Pi(s)$,
 and the width-like function,
$Im\Pi(s)$, for the I=1 channel. The crossing point of the running mass with
$s$ gives the $90^\circ$  mass of the \az .
 The magnitude of the \KK\ component in the \az\ is determined by
$-\frac d{ds}Re\Pi(s)$ which is large in the resonance region just
below the \KK\ threshold. These functions fix the
PWA of eq.(1) and Fig.~3. In Fig. 5 the running mass and width-like function
for the strange channel are shown. These fix the shape of the $K\pi$ phase
shift and absorption parameters in Fig. 1. 

Four out of our six parameters are fixed by the $K\pi$ data leaving only
$m_s-m_u=100$ MeV to "predict" the \az\ structure Fig.~3, and the
parameter $\beta$  to get the $\pi\pi$ phase shift right 
above 1 GeV/c. 
 One could discard the $\beta$
parameter if one also included the next group of important
thresholds or pseudoscalar ($0^{-+}$) -axial ($1^{+-}$)
 thresholds, since then the $K\bar K_{1B}+c.c.$
thresholds give a very similar contribution to the mass matrix as $\eta\eta'$.
As can be seen from Figs.~1-3 the model gives a good
description of the relevant  data.

\begin{table}[t] 
\caption{Resonances in the S-wave  $ PP \to PP $ amplitudes~$^1$ %\cite{NAT}.
The first resonance is the \sig\
which we name here $f_0(\approx 500)$. The two following are both
manifestations of the same $s\bar s$ state. The \fz\ and \az\ have no
approximate Breit Wigner-like description, and the $\Gamma_{BW}$ given for
\az\ is rather the peak width. 
The mixing angle $\delta_S $ for the
$f_0(\approx 500)$ or $\sigma$  is with respect to
$u\bar u +d\bar d$, while for the two heavier
$f_0$'s it is with respect to $s\bar s$. }
\vspace{ 0.5cm} 
\begin{center}
\begin{tabular}{|c|c|c|c|l|}
\hline
resonance&$m_{BW}$&$\Gamma_{BW}$&$\delta_{S,BW}$&Comment\\
\hline
$f_0(\approx 500)$&860&880&$(-9+i8.5)^\circ$&The $\sigma$ meson.\\
$f_0(980)$&-&-&-&First near \ss\ state\\
$f_0(1300)$&1186&360&$(-32+i1)^\circ$&Second
near \ss\ state\\
$K_0^*(1430)$&1349&498&-&The $s\bar d$ state\\
$a_0(980)$&987&$\approx$100&-&First I=1 state\\
\hline
\end{tabular}
\vspace{0.5cm}\end{center}
%\vspace{-1.5cm}\clearpage
\end{table}

\begin{table}[t] \caption{
The pole positions of the same resonances as in Table 1.
The last entry is an image pole of the
\az , which in an improved fit could represent the $a_0(1450)$.
The \fzz\ and \Kz\ poles appear simultaneously
on two sheets since the $\eta\eta$ and the $K\eta$ couplings, respectively,
 nearly vanish. The mixing angle $\delta_S $ for the
$f_0(\approx 500)$ or $\sigma$  is with respect to
$u\bar u +d\bar d$, while for the two heavier
$f_0$'s it is with respect to $s\bar s$. }
\vspace{ 0.5cm} 
\begin{center}
\begin{tabular}{|c|c|c|c|c|c|}
\hline
resonance&$s_{\rm pole}^{1/2}$& $[{\rm Re}s_{pole}]^{1/2}$&
$\frac{-{\rm Im}\ s_{pole}}{m_{pole}}$&$\delta_{S,pole}$&Sheet\\
\hline
$f_0(\approx 500)   $&$ 470-i250$& 397& 590 &$(-3.4+i1.5)^\circ $&II    \\
$f_0(980)   $&$1006-i17 $&1006& 34  &$(0.4+i39)^\circ   $&II    \\
$f_0(1300)  $&$1214-i168$&1202& 338 &$(-36+i2)^\circ    $&III,V \\
$K_0^*(1430)$&$1450-i160$&1441& 320 & -                  &II,III\\
$a_0(980)   $&$1094-i145$&1084& 270 & -                  &II    \\
$a_0(1450)? $&$1592-i284$&1566& 578 & -                  &III   \\
\hline
\end{tabular}
\vspace{0.5cm}\end{center}
%\vspace{-1.5cm}\clearpage
\end{table}

In Ref.~\cite{NAT} we looked for only those 
 four poles which are nearest to the
physical region, and which could complete a multiplet. 
These  were given in Ref.~\cite{NAT}: 
the \fz , \fzz , \az\ and \Kz .   We found parameters for these close to the
conventional lightest scalars in the 1994 PDG tables. 

\section{One $q\bar q$ pole can give rise to two resonances}

However, in addition there are other image poles~\cite{morgan}, usually
located far from the physical region.
As explained in Ref.~\cite{NAT2}, and below,
 some of these can come so close to
the physical region that they make new resonances.     
And, in fact, there are 
more than four physical poles  with different isospin,
in the output spectrum of our model, although only four bare states
are put in!
In Table 2 we list the significant  pole positions.

All these poles are manifestations of {\it the same nonet}~\cite{NAT2}.
The \fz\ and the \fzz\ turn out to be  two 
manifestations of the same \ss\ state.  
Fig. 6 shows how this can come about for the $s\bar s $ channel.
There can be two crossings with the running mass. Similarily the \az\ and the 
\azz\ are  likely to be two manifestations of the $u\bar d$ state. 

\section{The light $\sigma$ resonance}

A  light scalar-isoscalar meson (the \sig ), with a mass of
 twice the constituent  $u,d$ quark mass coupling strongly to
$\pi\pi$ is of importance in most models for spontaneous breaking
of chiral symmetry, and for our understanding of all hadron masses.
 Thus
most of the nucleon mass is believed to be generated by its coupling to
the $\sigma$, which acts like an effective Higgs-like 
 boson for the hadron spectrum.  However,  the lightest well established
mesons in the 1994 Review of Particle Properties 
with the quantum numbers of the $\sigma$, the \fz\ and \fzz\, did
not have the right properties. They are both too narrow, \fz\ couples mainly
to $K\bar K$, and \fzz\ is too heavy.

The important pole in \uu\ turns out to be  the the first pole in Table 2,
which is  the long sought for \sig =$f_0(\approx 500)$.
It gives rise to a very broad 
Breit-Wigner-like background, dominating 
 $\pi\pi$ amplitudes below 900 MeV.
It has the right mass and width and large $\pi\pi$
coupling as predicted by the \sig\ model.

The existence of this meson becomes evident if one studies the \uu\ channel
separately. This can be done within the model, perserving unitarity and
analyticity, by sending the $s$ quark (and $K$, $\eta$ etc.) mass
 to infinity. Thereby one  
eliminates the influence from \ss\ and \KK\ channels, which perturb $\pi\pi$
scattering very little through mixing below 900 MeV. 
The \uu\ channel is then seen to be dominated by the sigma below 900 MeV. 

Isgur and Speth~\cite{Isgur} have criticised this result 
claiming that crossed channel
exchanges, in particular $\rho$ exchange, is important. Our reply~\cite{Isgur}
to this is that because of the well known result from dual models, 
that a sum
of $s$-channel resonances also describe $t$-channel phenomena,
there is no inconsistency. A resonance amplitude is always a {\it product} of
a resonance pole and crossed channel singularities. In the model
of Ref.~$^{1,2}$ the crossed channel singularities were, in principle, included
through 
the form factor $F(s)$. Improvements to the model can be done by allowing
for a more complicated analytic form for $F(s)$, which then furthermore can be
constrained by data on exotic channels like $\pi^+\pi^+$ and $K^+\pi^+$
scattering.    

Recently we became aware of three references~\cite{sigmas},
which in addition to
those given in Ref.~$^2$ also support the existence of a light
$\sigma$ resonance, although within more limited models.
 The evidence for the $\sigma$ is thus mounting, and the
PDG tables of 1996 have now included it. Clearly this resonance is very
important for the understanding of the hadron spectrum as a whole. 

\section{Concluding remarks}

One could argue that the two states \fz\ and \az\ are a kind of
$K\bar K$ bound states (c.f. Ref.~\cite{wein}), 
since these have a large component of $K\bar K$
in their wave functions. However, the dynamics of these states is quite 
different from that of normal two-hadron bound states.
If one wants to consider them as \KK\ bound states,
it is the $K\bar K \to s\bar s \to K\bar K$ interaction which  creates their
binding energy. Thus, although they may spend most of their 
time as $K\bar K$ they 
owe their existence to the \ss\ state. Therefore it is more natural to consider
the \fz\ and \fzz\ as two manifestations of the same \ss\ state. 

%\setcounter{secnumdepth}{0} %this ensures that there are no section numbers

%\eject

\twocolumn[\hsize\textwidth\columnwidth\hsize\csname @twocolumnfalse\endcsname
]
\null\vfill
\begin{figure}
\psfig{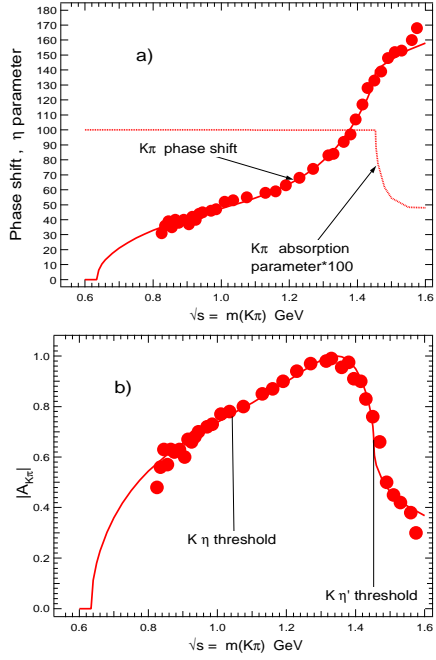}
\caption{ (a) The
$K\pi$ S-wave
phase shift and (b) the magnitude of the $K\pi$
partial wave amplitude compared with the model predictions, which
fix 4 ($\gamma$, $m_0+m_s$, $k_0$ and $s_{A,K\pi}$) of the 6
parameters.}
\end{figure}
\vfill
\eject
\null\vfill
\begin{figure}
\psfig{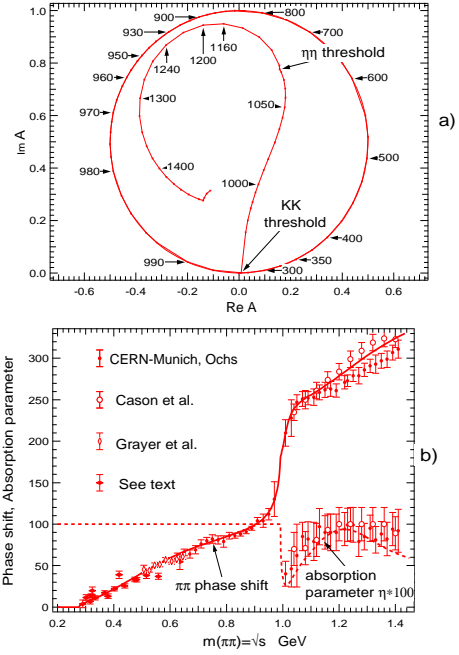}
\caption{ (a) The $\pi\pi$ Argand diagram and (b) phase shift predictions
are compared with data. Note that most of the parameters were fixed by the
data in Fig.1. For more details see Ref.~$^{1,2}$. }
\end{figure}
\vfill

\begin{figure}
\psfig{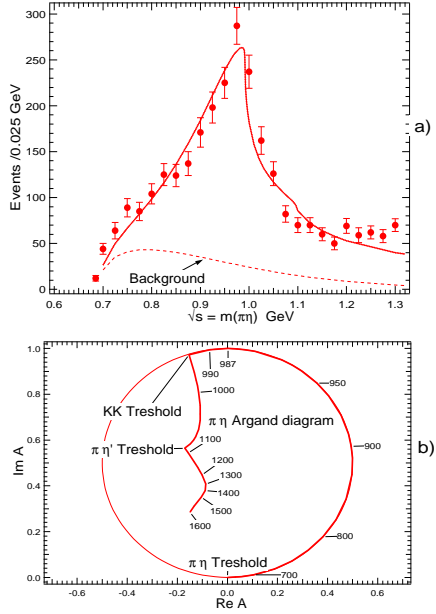}
\caption{ (a) The $a_0(980)$ peak compared with model prediction and (b) the 
predicted $\pi\eta$ Argand diagram}
\end{figure}

\begin{figure}
\psfig{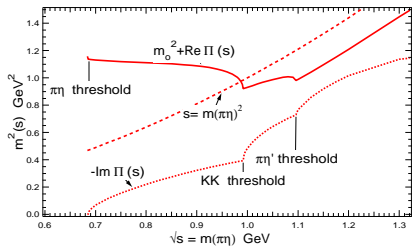}
\caption{ The running mass $m_0+{\rm Re}\Pi(s)$ and Im$\Pi (s)$ of the $a_0(980)$.
The strongly dropping running mass at 
the $a_0(980)$ position, below the $K\bar K$ 
threshold contributes to the narrow shape of the peak in Fig. 3a.}
\end{figure}
                                  
\begin{figure}
\psfig{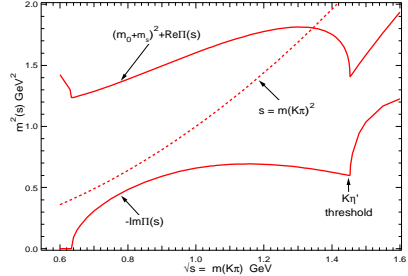}
\caption{The running mass and width-like function Im$\Pi(s)$ for the
$K^*_0(1430)$. The crossing of $s$ with the running mass gives the 90$^\circ$
phase shift mass, which roughly corresponds to a naive Breit-Wigner mass,
where the running mass is put constant.  }
\end{figure}
\vfill
\begin{figure}
\psfig{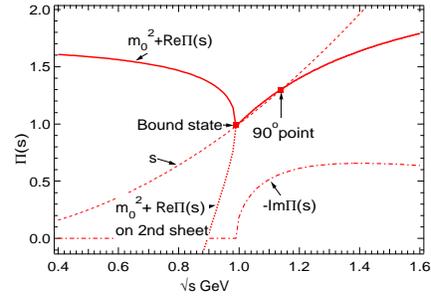}
\caption{ (a) Although the model has only one bare $s\bar s $  
resonance, when unitarized it can give rise to two crossings with 
the running mass in the $s\bar s - K\bar K$ channels. This means 
the $s\bar s$ state can manifest itself in
 two physical  resonances, one at threshold and
one near 1200 MeV (See Ref.~$^2$ for details) as in this figure. } 
\end{figure}

\end{document}